\documentclass[aps,preprint,tightenlines]{revtex4}
\usepackage{graphicx} 

\begin{document}

\title{Large $Q^2$ electrodisintegration of the deuteron}

\author{Misak M. Sargsian}
\affiliation{Department of Physics, Florida International University, Miami, FL 33199}

\begin{abstract}
The break up of the deuteron is studied at high $Q^2$ kinematics,
with main motivation to probe the deuteron at small internucleon distances. 
For this, two main issues are studied: 
electromagnetic interaction of the virtual photon with the bound nucleon 
and the strong interactions of the produced baryons in the final state of the reaction. 
Within virtual nucleon approximation we developed a new prescription to account for 
the  bound nucleon effects in the  electromagnetic interaction.
The  final state interaction at high $Q^2$ kinematics is calculated within generalized  
eikonal approximation~(GEA).  
Comparison with the first experimental data confirm GEA's early prediction that the 
rescattering is maximal at $\sim 70^0$ of recoil nucleon production  relative to 
the momentum of the virtual photon. Also  the forward recoil nucleon 
angles are best suited for studies of the deuteron at small distances.
\pacs{25.10.+s, 11.80.Fv, 25.30.Fj, 25.30.Rw}
\keywords      {Deuteron, High Energy Scattering}

\end{abstract}

\maketitle

\section{Introduction}

Deuteron represents an ideal micro-lab for studies of the structure of $NN$ interaction 
ranging from the intermediate to very small distances (see e.g. \cite{GG,gdpn,gdDD}.

The simplest reaction which could be used to investigate short-range $NN$ interactions 
in  nuclear targets is  the exclusive electrodisintegration of the deuteron in which 
large magnitudes of the  relative momentum of the  $pn$ system  in the ground state 
are  probed.   
Expectations that this can be achieved only at high-momentum transfer  
reactions\cite{FS8188,hnm,revsrc,MS01} was confirmed in a series of 
high energy experiments\cite{Day,Kim1,Kim2,Eip4,Eip5,Eip6} involving various nuclei.

Three experiments\cite{Ulmer,Kim3,Werner} have been performed recently using 
deuteron target at relatively  high~(up to 6~GeV) energy electron beam of the Jefferson Lab 
and more comprehensive experimental program will follow after the 12~GeV upgrade of the lab.
This  makes the development of theoretical approaches for the description of high 
$Q^2$ electro-nuclear processes involving deuteron a pressing issue.
Since mid 90's there were an intensive efforts in developing such theoretical approaches 
\cite{pdppn,Sabine1,edepn95,pdppnct,Sabine2,Ciofi1,Sabine,Sabine10,SabineOff_shell,Laget}.

In this work\cite{edepn} we study one particular process that is high $Q^2$ 
disintegration of the deuteron with probing large magnitudes of recoil 
nucleon momenta. The model is based on virtual nucleon approximation in 
which the struck nucleon is treated as an off-shell particle.   
The main theoretical  framework is based on the generalized eikonal 
approximation~(GEA)\cite{edepn95,gea,MS01,Tg1,Ciofi3,Ciofi4} which allows us to 
represent the reaction through the  set of covariant diagrams (Fig.1)
for which effective Feynman diagram rules can be defined. Because of the covariance 
of the diagrams  the virtualities involved in the scattering amplitudes are 
defined unambiguously.   This allows us to develop a self consistent approach 
for accounting for the binding effects in the high $Q^2$ electromagnetic interaction 
off the bound nucleon. The second important feature of GEA is that final-state 
interaction of produced two nucleons is calculated without requiring stationary 
approximation for the recoil nucleons - this is important feature for calculating 
processes in which the recoil nucleon is produced with large momenta.

\section{Main Assumptions of Virtual Nucleon Approximation}
First,  one considers only the $pn$  component of the deuteron,  
neglecting inelastic initial state transitions.
Since the deuteron is in a isosinglet state this will correspond to restricting  
the kinetic energy of  recoil nucleon to 
\begin{equation}
T_N < 2(m_\Delta-m_N) \sim (m_{N^*}-m) \sim 500~\mbox{MeV}
\label{asum1}
\end{equation}
where $m$,  $m_\Delta$ and $m_{N^*}$ are masses of the nucleon and  low-lying  
non-strange baryonic resonances.

Then, one neglects by the negative energy projection of the virtual nucleon propagator.
This can be justified if, 
\begin{equation}
M_d - \sqrt{m^2+ p^2}  > 0,
\label{asum2}
\end{equation}
where $M_d$ is the mass of the deuteron and $p$ is the relative momentum of the bound  $pn$ system.

The third assumption which is made in the calculation  is that at  large   $Q^2$~($> 1$~GeV$^2$)  
the interaction of virtual photon with exchanged mesons are a small correction and 
can  be neglected (see e.g. discussions in Ref.\cite{MS01,hnm}).
 
\begin{figure}[ht]
  \centering\includegraphics[scale=0.6]{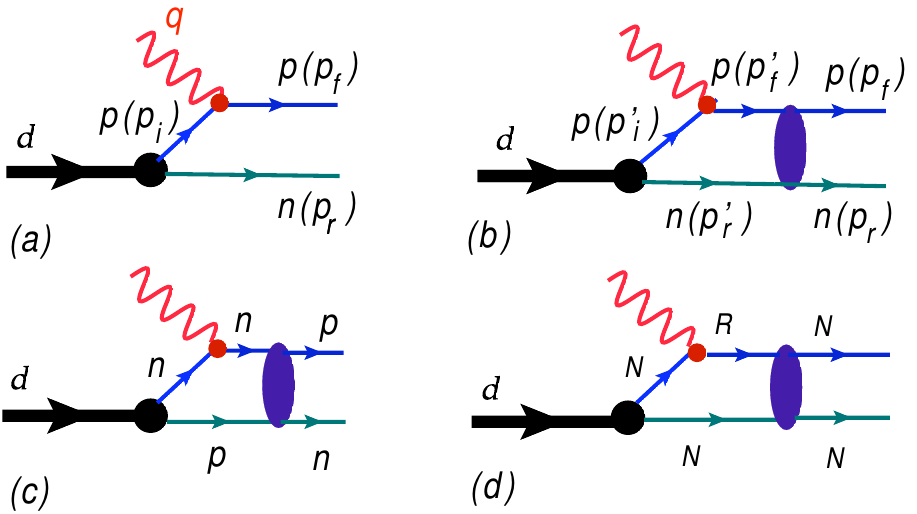}
  \caption{GEA diagrams}
\label{fig1}
\end{figure}

\section{Generalized Eikonal Approximation}

The assumptions discussed above allow us to restrict the consideration by the set of Feynman diagrams 
presented in Fig.(\ref{fig1}), which can be calculated based on the effective Feynman diagram rules 
derived  in GEA\cite{MS01}. This work does not include calculation of the diagram of Fig.1d 
which is currently in progress.

\medskip
\medskip 

\noindent{\bf Plane Wave Impulse Approximation Amplitude:}\\
Starting with the covariant form of the amplitude of Fig1.a, we calculate explicitly the 
off shell part of the electromagnetic amplitude through the one-shell positive 
energy  reduction of the  propagator of the spectator nucleon\cite{edepn}. 
This procedure  results to the PWIA amplitude\cite{edepn}:
\begin{equation}
\langle s_f,s_r \mid A_{0}^\mu\mid s_d\rangle  =  \sqrt{2}\sqrt{(2\pi)^3 2 E_r}\sum\limits_{s_i} J_{N}^\mu(s_f,p_f;s_i,p_i)
\Psi_d^{s_d}(s_i,p_i,s_r,p_r),
\label{A0_3}
\end{equation}
with 
\begin{equation}
J_{N}^\mu(s_f,p_f;s_i,p_i) = J_{N,on}^\mu(s_f,p_f;s_i,p_i)+J_{N,off}^\mu(s_f,p_f;s_i,p_i).
\label{J_N}
\end{equation}
where $J_{N,on}^\mu$ is the covariant on-shell electromagnetic current of the nucleon and 
\begin{equation}
J_{N,off}^\mu(s_f,p_f;s_i,p_i) = \bar u(p_f,s_f)\Gamma^\mu_{\gamma^*N} \gamma^0 u(p_i,s_i) \frac {E_i^{off} - E_i^{on}}{2m},
\label{J_off}
\end{equation}
in which  $E_i^{off} = M_d - E_i^{on}$ and $E_i^{on} = \sqrt{m^2 + p_i^2}$, $\vec p_i = - \vec p_r$.
Note that the total current in Eq.(\ref{J_N}) is conserved since it is derived from 
the gauge invariant amplitude and no additional conditions  are needed to restore 
the  current conservation.  

\medskip
\medskip 

\noindent{\bf Forward Elastic and Charge Interchange Final State Interaction Amplitudes:}\\
Applying the effective Feynman diagram rules to the diagrams of Fig.(\ref{fig1})b,c and projecting  
(similar to PWIA) the propagator of the spectator nucleon to its  positive energy solution for 
the forward rescattering amplitude we obtain:
\begin{eqnarray}
& & \langle s_f,s_r \mid A_{1}^\mu\mid s_d\rangle  =     {i \sqrt{2}(2\pi)^{3\over 2}\over 4} \sum\limits_{s^\prime_f,s^\prime_r,s_i} 
\int {d^2p_r^\prime \over  (2\pi)^2}  \frac{\sqrt{2\tilde E^\prime_r}\sqrt{s(s-4m^2)}} {2\tilde E^\prime_r |q|}  \times \nonumber \\
& &  \ \ \ \ \ \  \langle p_f,s_f;p_r,s_r\mid f^{NN,on}(t,s)\mid \tilde p^\prime_r,s^\prime_r; \tilde p^\prime_f,s^\prime_f\rangle 
\cdot  J_{N}^\mu(s^\prime_f,p^\prime_f;s_i,\tilde p^\prime_i) \cdot  \Psi_d^{s_d}(s_i,\tilde p^\prime_i,s^\prime_r,\tilde p^\prime_r) 
\nonumber  \\
& & -  {\sqrt{2}(2\pi)^{3\over 2}\over 2} \sum\limits_{s^\prime_f,s^\prime_r,s_i} {\cal P}\int {dp^\prime_{r,z}\over 2\pi} 
\int {d^2p_r^\prime \over  (2\pi)^2}  \frac{\sqrt{2E^\prime_r}\sqrt{s(s-4m^2)}} {2E^\prime_r |{\bf q}|} 
\times \nonumber \\
 & & \ \ \ \ \ \  {\langle p_f,s_f;p_r,s_r\mid f^{NN,off}(t,s)\mid p^\prime_r,s^\prime_r;p^\prime_f,s^\prime_f\rangle
\over  p^\prime_{r,z}- \tilde p^\prime_{r,z} }  J_{N}^\mu(s^\prime_f,p^\prime_f;s_i,p^\prime_i)
\cdot  \Psi_d^{s_d}(s_i,p^\prime_i,s^\prime_r,p^\prime_r),
\label{a1_fsi}
\end{eqnarray}
where $\tilde p^\prime_r = (p_{r,z}-\Delta, p^\prime_{r,\perp})$, $\tilde E^\prime_r = \sqrt{m^2 + \tilde p^{\prime 2}_r}$, 
$\tilde p^\prime_i = p_d - \tilde p^\prime_r$ and $\tilde p^\prime_f = \tilde p^\prime_i + q$.

In high energy limit in which  the helicity conservation of small angle NN scattering is rather 
well established the  on-shell amplitude  is  predominantly imaginary and can be parameterized in the form
\begin{equation}
 \langle p_f,s_f;p_r,s_r\mid f^{NN,on}(t,s)\mid \tilde p^\prime_r,s^\prime_r; \tilde p^\prime_f,s^\prime_f\rangle  
= \sigma_{tot}^{pn}(i + \alpha) e^{{B\over 2}t} \delta_{s_f,s^\prime_f}\delta_{s_r,s^\prime_r},
\label{fnn_on}
\end{equation}
where $\sigma^{pn}_{tot}(s)$,  $B(s)$ and $\alpha(s)$ are found from fitting of experimental data on elastic 
$pn\rightarrow pn$ scattering. 
For the half-off-shell part of the $f^{NN,off}$ amplitude we use the following parameterization:
\begin{equation}
f^{NN,off} = f^{NN,on} e^{B(m_{off}^2 -m^2)},
\label{fnn_off}
\end{equation}
where $m^2_{off} = (M_d-E^\prime_{r} + q_0 )^2 - (p^\prime_r+q)^2$.

For the charge-exchange final state interaction amplitude~(Fig.1c) the derivation is similar 
and it is expressed through the charge-exchange $pn\rightarrow np$ scattering amplitude.\\

\begin{figure}[th]
\includegraphics[scale=0.3]{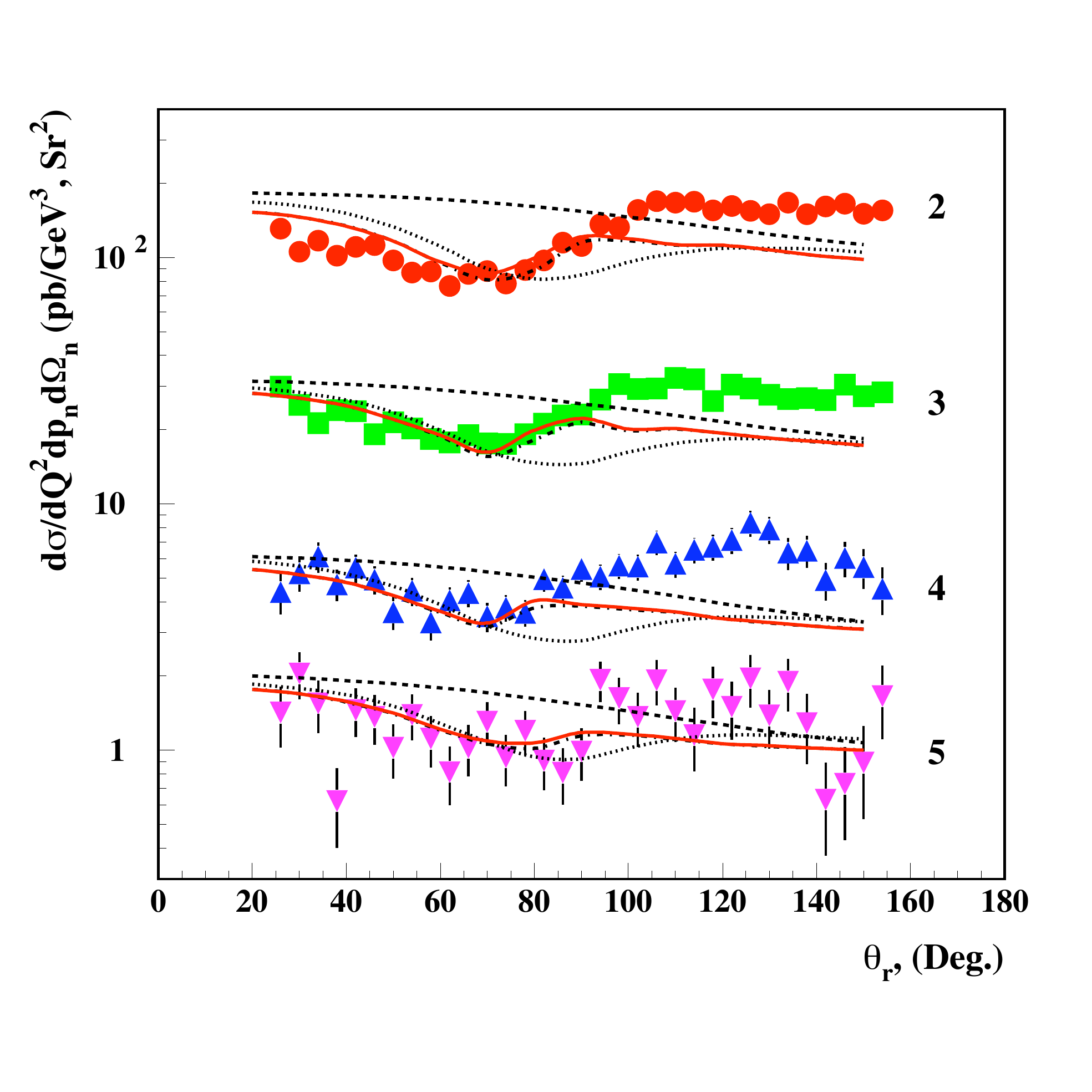}
\includegraphics[scale=0.3]{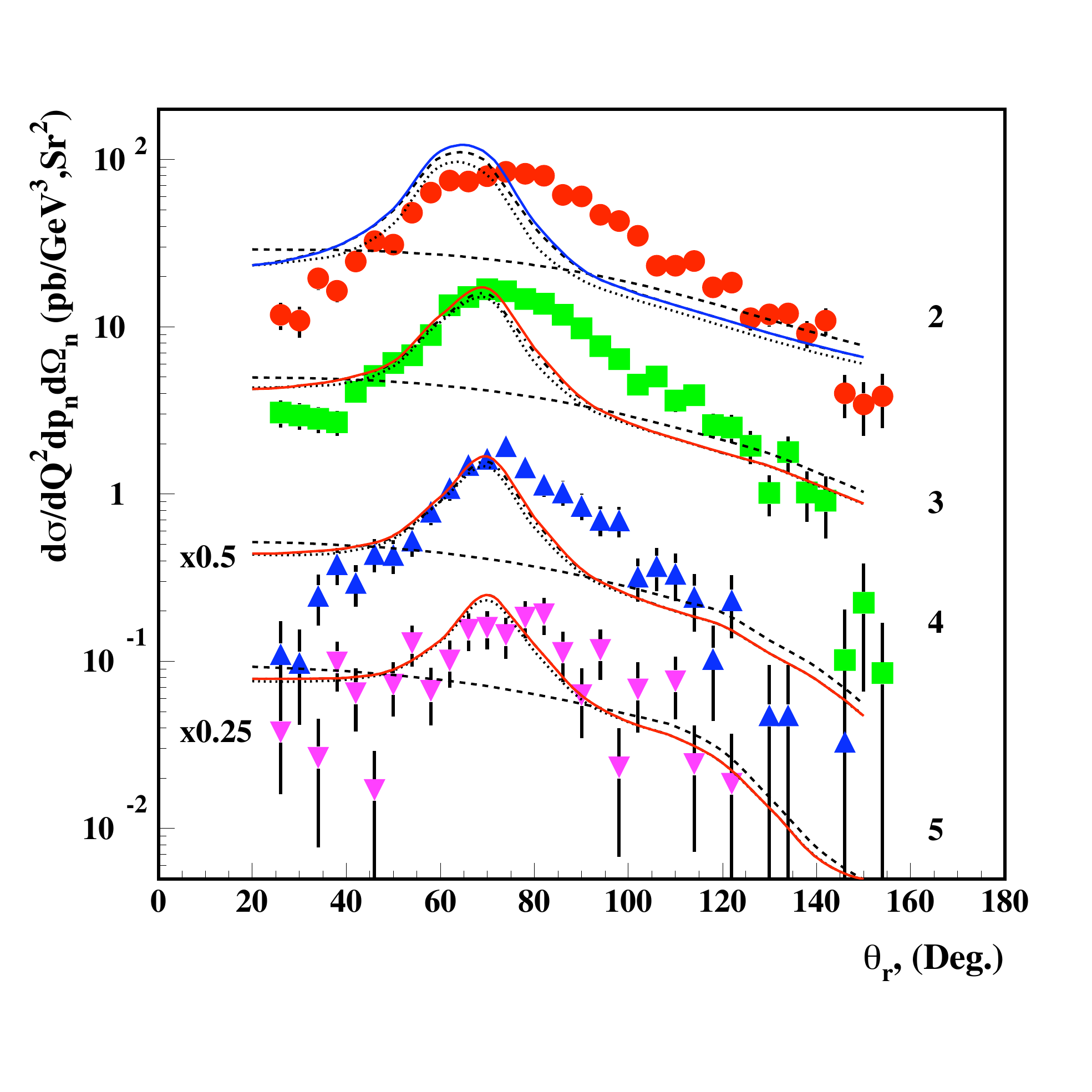}

\caption{Dependence of the differential  cross section on the direction of 
the recoil neutron momentum. 
The data are from Ref.\cite{Kim3}.
Dashed line - PWIA calculation, dotted line - PWIA+ pole term of forward FSI,  
dash-dotted line  -  PWIA+forward FSI,  
solid line - PWIA + forward and charge exchange FSI.  The momenta of the recoil neutron 
in (a)  and (b) are  restricted to $200 < p_r < 300$~MeV/c and  $400 < p_r < 600$~MeV/c 
respectively.
The labels 2, 3, 4 and 5 correspond to the following values of 
$Q^2 = 2\pm 0.25;  3\pm 0.5; 4\pm 0.5; 5\pm 0.5$~GeV$^2$.  The data sets and calculations for 
``4'' and ``5'' in (b) are multiplied by  0.5 and 0.25 respectively}
\label{fig2}
\end{figure}


\noindent{\bf Deuteron Wave Function:}\\
To fix the normalization of the deuteron wave function we use the fact that 
the deuteron elastic charge form-factor $G_C\rightarrow 1$ at $Q^2\rightarrow 0$.
Using above condition in the elastic deuteron scattering amplitude one can 
relate the deuteron wave function in virtual nucleon approximation to the 
nonrelativistic deuteron wave function as follows\cite{edepn}  
\begin{equation}
\Psi_d(p) = \Psi^{NR}_d(p) {M_d\over 2(M_d - \sqrt{m^2+p^2})}. 
\label{wfmod}
\end{equation}
This relation provides a smooth transition to the nonrelativistic wave function $\Psi^{NR}$ in 
the small momentum limit.

\section{Comparison with Experimental Data and Some Conclusions}

In the last few years  three experiments\cite{Ulmer,Kim3,Werner} produced the first data
at relatively large $Q^2$ kinematics.
The experiment of Ref.\cite{Kim3}) covered the  highest to date $Q^2$ range from 
$2-5$~GeV$^2$. Comparison with these experimental data are given in Fig.\ref{fig2}
which allow us to make the following conclusions: 

- the angular distribution clearly exhibits   an eikonal feature, with the minimum (Fig.\ref{fig2}(a)) 
or maximum (Fig.\ref{fig2}(b))  at transverse kinematics 
due to the  final state interaction.  The most important result is that the maximum of FSI is at
recoil angles of $70^0$   in agreement with the GEA prediction of Ref.\cite{gea}. 
Note that the  conventional  Glauber theory predicted $90^0$ for the FSI maximum.

- The disagreement of the calculation with the data at  $\theta_r > 70^0$ 
appears to be due to the isobar contribution at the 
intermediate state of the reaction. This region corresponds to $x<1$ and it is   
kinematically  closer to the threshold  of $\Delta$-isobar 
electroproduction. The comparisons also indicate 
that the relative strength of the  $\Delta$-isobar contribution diminishes with an increase of 
$Q^2$ and at neutron production angles $\theta_r\rightarrow 180^0$.  

- The forward direction of the recoil nucleon momentum, being far from the $\Delta$-isobar threshold, 
exhibits a relatively small contribution due to FSI.  This indicates that the forward recoil angle 
region is best suited for studies of  PWIA properties of the reaction such as the off-shell 
electromagnetic current and deuteron wave function.

\medskip
\medskip

These comparisons clearly show that the forward angles of spectator nucleon  production is 
best suited for isolating PWIA scattering off the virtual nucleon. Therefore this kinematic 
region provides the most optimal condition for probe and studying the  $NN$ interaction at 
short internucleon separations.


\medskip

\noindent{\bf Acknowledgments:}\\
This work is supported by the  U.S. Department of Energy Grant 
under Contract DE-FG02-01ER41172.

\end{document}